\documentclass[prl,
amsmath,amssymb,
superscriptaddress,
longbibliography,
twocolumn]{revtex4-1}

\usepackage{graphicx}
\usepackage{dcolumn}
\usepackage{bm}
\usepackage{color}
\usepackage{hyperref}

\begin{document}

\newcommand{\fix}[1]{\noindent  {\colorbox{red}{\color{white}
FIX:}  \color{red} #1\normalcolor}}

\title{Designing morphology of separated phases in multicomponent liquid mixtures}
\author{Sheng Mao}
\thanks{Equal contributions}
\affiliation{Department of Mechanics and Engineering Science, BIC-ESAT, College of Engineering, Peking University, Beijing 100871, People's Republic of China}
\affiliation{Department of Mechanical and Aerospace Engineering, Princeton University, Princeton, New Jersey 08544, USA}

\author{Milena S. Chakraverti-Wuerthwein}
\thanks{Equal contributions}
\affiliation{Department of Physics, Princeton University, Princeton, New Jersey 08544, USA}

\author{Hunter Gaudio}
\affiliation{Department of Mechanical and Aerospace Engineering, Princeton University, Princeton, New Jersey 08544, USA}
\affiliation{Department of Mechanical Engineering, Villanova University, Villanova, Pennsylvania, 19085, USA}

\author{Andrej Ko\v{s}mrlj}
\email[]{andrej@princeton.edu}
\affiliation{Department of Mechanical and Aerospace Engineering, Princeton University, Princeton, New Jersey 08544, USA}
\affiliation{Princeton Institute for the Science and Technology of Materials (PRISM),
Princeton University, Princeton, New Jersey 08544, USA}

\date{\today}

\begin{abstract}
Phase separation of multicomponent liquid mixtures plays an integral part in many processes ranging from industry to cellular biology. In many cases the morphology of coexisting phases is crucially linked to the function of the separated mixture, yet it is unclear what determines morphology when multiple phases are present. We developed a graph theory approach to predict the topology of coexisting phases from a given set of surface energies (forward problem), enumerate all topologically distinct morphologies, and reverse engineer conditions for surface energies that produce the target morphology (inverse problem). 
\end{abstract}


\maketitle

Phase separation and multi-phase coexistence are ubiquitous ranging from the simple demixing of water and oil to more sophisticated industrial processes related to medicine, food, cosmetics, energy, environment,  etc.~\cite{lohse2020physicochemical} Phase separation and multi-phase coexistence also occur in nature, where they give rise to structural colors in birds~\cite{dufresne2009self,parnell2015spatially,burg2018self} and produce a plethora of intracelullar condensates~\cite{hyman2014liquid,Shin2017,berry2018physical,choi2020physical}.

Coexisting liquid phases can adopt a variety of different morphologies~\cite{utada2005monodisperse, roh2006triphasic,shah2008designer,choi2013one, zarzar2015dynamically, nagelberg2017reconfigurable,moerman2018emulsion}, which are often directly linked to some function, e.g. the nested morphology of separated phases can assist with drug delivery~\cite{haase2014tailoring} and with the biogenesis of ribosomes inside cell nuclei~\cite{feric2016}, while the tunable morphologies of multi-phase droplets can serve as micro-lenses with tunable focal length~\cite{nagelberg2017reconfigurable}.
The control of morphology of separated liquid phases could open the avenue for new applications, but we currently lack tools for designing the morphology of more than three coexisting phases. In this Letter we make an important step in this direction.

The phase separation process is rooted in thermodynamics and the main principles have been known since Gibbs~\cite{gibbs1878art}. More recently these arguments have been extended to multicomponent systems and several tools have been developed that enable predicting the number of coexisting phases, their compositions and volume fractions, and surface energies between them~\cite{panagiotopoulos1987direct,panagiotopoulos1988phase,frenkel2001understanding,cool2010gibbs,koukkari2011gibbs, wolff2011,mao2019phase}. While the minimization of the bulk free energy determines the number of coexisting phases, their compositions and volume fractions, the minimization of surface energies determines how these phases arrange in space. (Here, we neglect buoyancy effects, hydrodynamics, and chemical reactions, which can also affect morphology~\cite{koga1991spinodal, berry2018physical,lohse2020physicochemical}.) 

The focus of this Letter is to explain how surface energies determine the topology of separated liquid phases, but we also briefly comment how volume fractions affect the geometry of separated phases. The topology of separated phases can be represented with a connectivity graph. We show how to use graph theory to predict the topology of separated phases from a given set of surface energies (forward problem), enumerate all topologically distinct morphologies, and reverse engineer conditions for surface energies that produce the target morphology (inverse problem).

The graph theory approach presented below is general and can be applied to any model system. Here, we use the Flory-Huggins~\cite{huggins1941,flory1942} model of regular solutions together with a Cahn–-Hilliard approach for kinetics and interfacial energies~\cite{cahn1958} to validate predictions from the graph theory approach for $N_p=3$, $4$, and $5$ coexisting phases in 3D. The free energy density $f$  of the mixture with $N_c$ different components is written as~\cite{hoyt1990continuum, mao2019phase,berry2018physical}
\begin{equation}
\frac{f}{c RT} = \sum_{i=1}^N \phi_i \ln \phi_i 
+ \frac{1}{2}\sum_{i, j=1}^N \chi_{ij} \phi_i \phi_j - \frac{\lambda^2}{2} \sum_{i,j=1}^N \chi_{ij} \nabla \phi_i \nabla \phi_j,
\label{eq:fh}
\end{equation}
where $c$ is the total concentration of the mixture, $R$ the gas constant, $T$ the temperature, $\phi_i$ the volume fraction of the component $i$ with $\sum_{i} \phi_i = 1$, $\chi_{ij}$ the interaction parameter between components $i$ and $j$ with $\chi_{ii} = 0$, and $\lambda$ is the characteristic width of the interface. In the above Eq.~(\ref{eq:fh}) the three terms describe the entropy of mixing, the interaction energy, and the interfacial energy~\footnote{Note that the negative sign for the interfacial energy is due to the incompressibility $\sum_{i} \nabla \phi_i = 0$ as discussed in ~\cite{mao2019phase}}. The volume fractions evolve as
\begin{equation}
\frac{\partial \phi_i}{\partial t} = D \nabla \cdot \big[\phi_i \sum_j \left(\delta_{ij}-\phi_j\right) \nabla \tilde \mu_j \big],
\end{equation}
where $D$ is the diffusion coefficient~\footnote{We assume that all components have the same diffusion coefficient $D$, but the respective mobilities are different and composition-dependent. See Supplementary Material and~\cite{mao2019phase} for details.}, $\delta_{ij}$ the Kronecker delta, and $\tilde \mu_j=1+\ln \phi_j + \sum_k \chi_{jk} (1+\lambda^2 \nabla^2)\phi_k$ are the dimensionless chemical potentials. Here, we also assume that the interaction parameters $\chi_{ij}$ are sufficiently large, such that the mixture separates into $N_p=N_c$ distinct phases via spinodal decomposition, where each of the phases $I$ is enriched with the component $i$~\cite{mao2019phase}, and the volume fractions of separated phases are approximately equal to the average volume fractions $\{\overline \phi_i\}$ of components. In this limit, the surface energies can be estimated as  $\gamma_{IJ}\approx (\pi c \lambda RT/4) \chi_{ij}$.
The details of the simulations are provided in the Supplemental Material and in Ref.~\cite{mao2019phase}.

\begin{figure}[t!]
\includegraphics[scale=1]{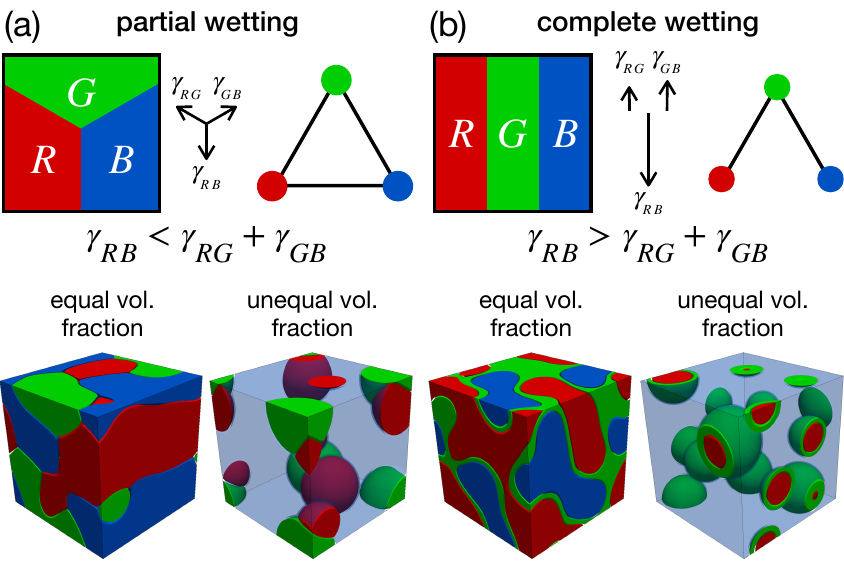}
\caption{\label{fig1} 
Morphologies of three coexisting phases $R$ (red), $G$ (green), and $B$ (blue) are determined by the magnitudes of surface tensions ($\gamma_{RB} \geq \gamma_{RG}\geq \gamma_{BG}>0$) and volume fractions. 
First row: schematics of local arrangements of phases and corresponding graph representations for (a)~partial wetting ($\gamma_{RB}<\gamma_{RG}+\gamma_{GB}$) with stable triple junctions due to the force balance of surface tensions, and (b)~complete wetting ($\gamma_{RB}>\gamma_{RG}+\gamma_{GB}$) with unstable triple junctions due to the force imbalance of surface tensions.
Second row: representative simulation snapshots at $10^6$ timesteps (see \href{http://www.princeton.edu/~akosmrlj/papers/phase_separation_design/videos/}{Video~S1} for the time evolution~\cite{video}).
The blue phase is semi-transparent for the snapshots with unequal volume fractions. The simulation parameters are given in Table~S1~\cite{SI}.}
\end{figure}

To introduce relevant concepts, we first discuss the morphology of three coexisting phases R (red), G (green), and B (blue), with surface energies $\gamma_{RB}\geq \gamma_{RG}\geq \gamma_{GB}>0$. When surface energies satisfy the triangle inequality ($\gamma_{RB}< \gamma_{RG}+ \gamma_{GB}$), the phases partially wet each other. Triple junctions, where three phases meet, are stable (see Fig.~\ref{fig1}a) and they persist during the coarsening (see \href{http://www.princeton.edu/~akosmrlj/papers/phase_separation_design/videos/}{Video~S1}~\cite{video}). The equilibrium angles between different phases can be obtained from the force-balance of surface tensions, which is known as the Neumann construction~\cite{de2013capillarity}. In contrast, when surface energies do not satisfy the triangle inequality ($\gamma_{RB} > \gamma_{RG}+ \gamma_{GB}$), the phase $G$ completely wets the phases $R$ and $B$ to eliminate the high surface energy $\gamma_{RB}$ (see Fig.~\ref{fig1}b and \href{http://www.princeton.edu/~akosmrlj/papers/phase_separation_design/videos/}{Video~S1}~\cite{video}). Here, triple junctions are unstable, because surface tensions $\gamma_{RG}$ and $\gamma_{GB}$ cannot balance the high surface tension $\gamma_{RB}$ (Fig.~\ref{fig1}b).

The topology of separated phases can be represented with a connectivity graph, where vertices correspond to phases and edges connect phases that share a 2D interface. Note that phases that meet only at points or 1D lines are disconnected in the graph representation.
The fully connected graph describes the case with partial wetting, where all phases are in contact with each other (Fig.~\ref{fig1}a), while the graph with a missing edge corresponds to the case with complete wetting (Fig.~\ref{fig1}b). Note that the topology of separated phases is fully determined by surface tensions, while the geometry of separated phases also depends on the volume fractions of phases (Fig.~\ref{fig1}). Phases percolate through the whole space, when their volume fractions exceed the percolation threshold ($\approx 0.34$ in 3D~\cite{stauffer1994introduction}), but otherwise they break into droplets to minimize the surface energy, which is known as the Plateau-Rayleigh instability~\cite{de2013capillarity, eggers1997nonlinear}.

The information presented above for mixtures with three coexisting phases can be used to infer the behavior of mixtures with $N_p>3$ coexisting phases. For any model system with $N_c$ components, the first step is to predict the number $N_p$ of coexisting phases, their compositions and volume fractions, and surface energies $\{\gamma_{IJ}\}$ between them by using the tools described in Refs.~\cite{panagiotopoulos1987direct,panagiotopoulos1988phase,frenkel2001understanding,cool2010gibbs,koukkari2011gibbs, wolff2011,mao2019phase}. For each of the $N_p \choose 3$ subsets of three phases $\{I,J,K\}$, the local arrangement of phases depends on the surface energies $\{\gamma_{IJ},\gamma_{IK},\gamma_{JK}\}$ and can be represented with triplet connectivity graphs (Fig.~\ref{fig1}). The fully connected graph corresponds to the partial wetting case with stable triple junctions $I-J-K$, where surface energies satisfy the triangle inequality ($\gamma_{IJ}<\gamma_{IK}+\gamma_{JK}$, $\gamma_{IK}<\gamma_{IJ}+\gamma_{JK}$, $\gamma_{JK}<\gamma_{IJ}+\gamma_{IK}$). The graph with a missing edge $I-J$  describes the case where the phase $K$ completely wets the phases $I$ and $J$ and surface energies satisfy the inequality $\gamma_{IJ}>\gamma_{IK}+\gamma_{JK}$. Analogously we can interpret the two other graphs with either a missing edge $I-K$ or a missing edge $J-K$.

The information from the triplet connectivity graphs for each of the $N_p \choose 3$ subsets of three phases can be used to construct the connectivity graph for the whole system with $N_p$ phases. Starting with a fully connected graph with $N_p$ vertices, we iterate over each of the $N_p \choose 3$ graphs and for each edge missing in the triplet graph, we remove the corresponding edge in the $N_p$ connectivity graph. This yields a connectivity graph that describes the topology of the mixture.

\begin{figure}[t!]
\includegraphics[scale=1]{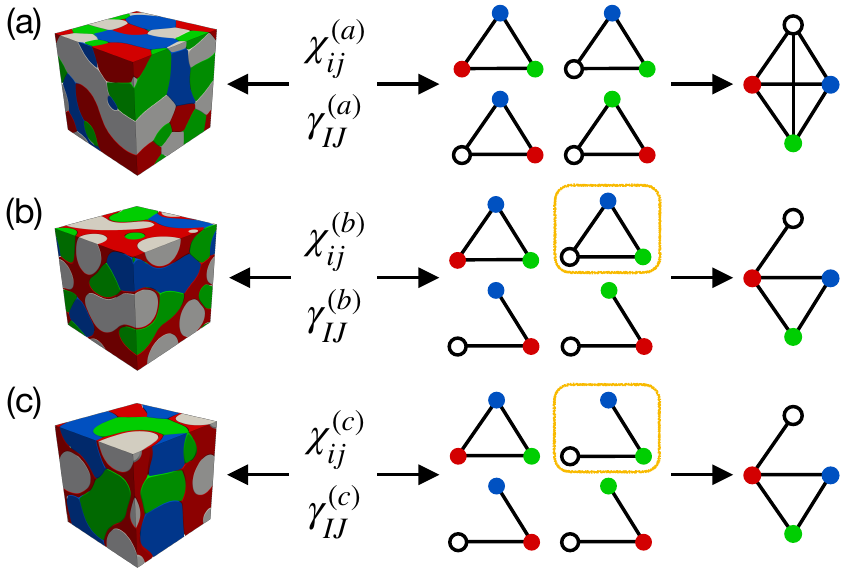}
\caption{\label{fig2}
Prediction of the topology of separated phases. From the set of interaction parameters $\{\chi_{ij}\}$ and average volume fractions $\{\overline \phi_i \}$ of components, we can predict the surface energies $\{\gamma_{IJ}\}$ of separated phases. These values are then used to produce the set of graphs of triplets of phases, from which we construct the connectivity graph describing the topology of separated phases (see text). These graphs are then compared to the topology of separated phases in simulation snapshots at $10^6$ timesteps on the left. (b,c) Different sets of graphs (yellow boxes indicate the difference) can produce the same topology of separated phases. The simulation parameters are given in Table~S1~\cite{SI}.
}
\end{figure}

Fig.~\ref{fig2} shows a few representative cases for mixtures with $N_p=4$ coexisting phases (red, green, blue, white), where ${4 \choose 3}=4$ graphs of triplets of phases are used to construct the connectivity graph with $4$ vertices that describes the topology of separated phases. When all $4$ graphs of triplets of phases are fully connected, then the connectivity graph with $4$ vertices is also fully connected (see Fig.~\ref{fig2}a). Distinct sets of triplet graphs can construct the same 4-component connectivity graph (Fig.~\ref{fig2}b,c). One such example can be seen in Fig.~\ref{fig2}b,c, where the graph is missing an edge between the white and blue and between the white and green phases because the red phase completely wets the white and blue and white and green phases. The different wetting condition (highlighted in Fig.~\ref{fig2}b,c) between the white, green, and blue phases in these distinct cases do not affect the final connectivity graph (or topology), but they affect the transient dynamics. For the case in Fig.~\ref{fig2}b, the white, green, and blue phases form stable triple junctions, which get broken once the red phase comes along and separates the white phase from the green and blue phases. In contrast, for the case in Fig.~\ref{fig2}c, the green phase completely wets the white and blue phases, but the presence of the red phase separates the green and white phases.

We also checked that the connectivity graphs accurately predict the topology of separated phases in simulations with $N_c=4$ components (see Fig.~\ref{fig2}), where the interaction parameters $\chi_{ij}\propto \gamma_{IJ}$ were chosen to be consistent with the set of inequalities for surface energies described by the $4$ graphs of triplets of phases. 

\begin{figure}[t!]
\includegraphics[scale=1]{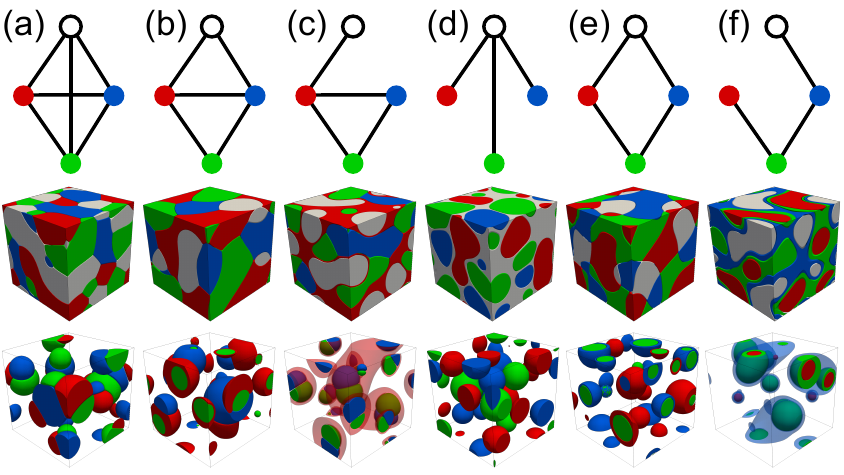}
\caption{\label{fig3} 
Graph representations and simulation snapshots at $10^6$ timesteps for all distinct topologies of $4$ coexisting phases with equal volume fractions (top) and non-equal volume fractions with transparent white phase (bottom). See \href{http://www.princeton.edu/~akosmrlj/papers/phase_separation_design/videos/}{Video~S2} for time evolution~\cite{video}. The simulation parameters are given in Table~S1~\cite{SI}.
}
\end{figure}

\begin{figure*}[t!]
\includegraphics[scale=1]{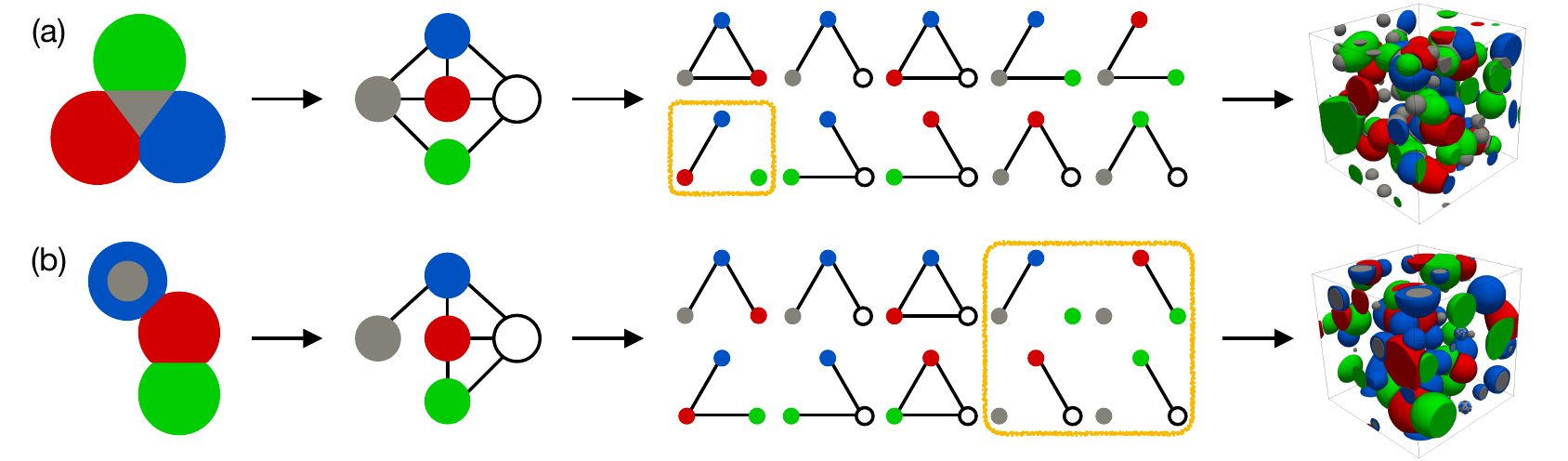}
\vspace{-0.4cm}
\caption{\label{fig4} 
Reverse engineering of target structures. To reverse engineer the model parameters for target structures, we first construct a connectivity graph, which is then divided into subgraphs of triplets of phases that are associated with inequalities of surface energies. The subgraphs highlighted with yellow boxes do not provide any constraints on surface energies.
The linear programming is used to find a set of surface energies that satisfy these inequalities (see text), which are then converted to interaction parameters $\chi_{ij}$. The average volume fractions $\{\overline \phi_i\}$ of components are chosen such that the volume fractions of separated phases are consistent with the target structure. The resulting simulation snapshots at $5\times 10^5$ timesteps  are shown on the right (see \href{http://www.princeton.edu/~akosmrlj/papers/phase_separation_design/videos/}{Video~S3} for time evolution~\cite{video}). The simulation parameters are given in Table~S1~\cite{SI}.
}
\end{figure*}

The representation of the topology of separated phases in terms of the connectivity graphs enables us to enumerate all topologically distinct morphologies, which correspond to all connected unlabelled graphs~\cite{harary2014graphical}. For $N_p=3$ phases there are two distinct graphs, which are shown in Fig.~\ref{fig1}. For $N_p=4$ phases there are $6$ distinct graphs (see Fig.~\ref{fig3}), which can all be realized by appropriately adjusting surface energies (as described below). Since some of the topologies can be obtained from multiple sets of graphs for triplets of phases (see Fig.~\ref{fig2}b,c), we systematically investigated all possibilities for the mixture with $N_p=4$ phases. 

First we generated all $4^{N_p \choose 3}=4^4=256$ sets of  ${N_p \choose 3}=4$ graphs of triplets of phases, where each graph can either be fully connected or is missing one of the 3 edges. Then we removed all duplicate sets of graphs that can be obtained by permutations of labels, resulting in $19$ distinct sets of graphs (see Figs.~S1 and S2~\cite{SI}). Each set of graphs of triplet phases corresponds to a set of inequalities for surface energies $\{\gamma_{IJ}\}$ as described above, which can have either infinite solutions or no solutions. We found that $6$ of the $19$ sets have no solutions (see Figs.~S2~\cite{SI}). To obtain representative values of interaction parameters $\{\chi_{ij}\}$ for the other $13$ sets (see Figs.~S1~\cite{SI}), we solved a linear programming problem by minimizing the sum $\sum_{ij} \chi_{ij}$ subject to the inequalities provided by the set of graphs, where we took into account that $\chi_{ij} \propto \gamma_{IJ}$.  To ensure that the inequalities were strictly enforced we added a small value of $\epsilon=0.2$-$0.5$ to each inequality, e.g. $\chi_{ij} \geq \epsilon + \chi_{ik}+\chi_{jk}$. Furthermore, we imposed additional constraints $\chi_{ij}\geq\chi_\text{min}=2$--$3$, where the value of $\chi_\text{min}$ has to be sufficiently large to ensure that the mixture actually separates into $4$ phases via spinodal decomposition~\cite{mao2019phase}.

This way we were able to obtain representative simulations for all $13$ distinct sets of graphs (see Fig.~S1~\cite{SI}) and the topologies of separated phases were consistent with predictions from the graph theory approach described above. These $13$ cases can be grouped in $6$ distinct topologies, which are shown in Fig.~\ref{fig3} (see \href{http://www.princeton.edu/~akosmrlj/papers/phase_separation_design/videos/}{Video~S2} for time evolution~\cite{video}), where we also show how changes in volume fraction of phases change the geometry, but not the topology of separated phases. (Note that the morphologies in 2D and 3D are  equivalent (Fig.~S3~\cite{SI})). 
Note that in Fig.~\ref{fig3}e we observed stable quadruple junctions, where all $4$ phases meet~\footnote{Note that for the snapshot in Fig.~\ref{fig3}e the green and the white phases meet at quadruple junctions, but this is not in conflict with the connectivity graphs, where the connected vertices correspond to phases that share a 2D interface.}. While quadruple junctions are typically energetically unstable, we show that for this case the conditions for surface tensions are such that they stabilize the quadruple junctions (see Fig.~S4~\cite{SI}). 

The number of distinct topologies (i.e. the number of connected unlabelled graphs) rapidly increases with the number $N_p$ of coexisting phases and scales as $e^{\alpha N_p^2}$, where $\alpha\sim0.3$~\cite{harary2014graphical}. It remains unclear whether all of them can actually be realized by appropriately tuning the values of $N_p \choose 2$ surface energies.

Finally, we also comment on how to reverse engineer model parameters to obtain target structures. Fig.~\ref{fig4} sketches the procedure for two target morphologies with $N_p=5$ coexisting phases. Starting from a target structure, we construct the connectivity graph, where vertices correspond to phases and edges connect phases that share a 2D interface. The connectivity graph can then be broken down into ${N_p \choose 3} = 10$ subgraphs for triplets of phases. Each subgraph with three edges (partial wetting) or two edges (complete wetting) can be directly translated to the inequalities for surface energies as described above. However, there could also be subgraphs with only one edge or no edges (highlighted with yellow boxes in Fig.~\ref{fig4}), which do not provide any restrictions on surface energies. For the case in Fig.~\ref{fig4}a the 9 subgraphs provide enough conditions on surface tensions to generate the target connectivity graph with $5$ vertices and no additional constraints are needed for the red-green-blue subgraph. However, for the case in Fig.~\ref{fig4}b the 6 subgraphs are not sufficient and we need to impose another restriction to ensure that the edge between the green and dark gray phases is removed, e.g. by requiring that the red ($R$) phase wets the green ($G$) and dark gray ($D$) phases ($\gamma_{GD}>\gamma_{RG}+\gamma_{RD}$). The set of surface energies can then be obtained by solving the linear programming problem subject to the inequalities imposed by the subgraphs and any other constraints that may be provided by the model or experimental system. The next step is to convert the values of surface energies to interaction parameters between components. This is in general a highly nontrivial inverse problem, but here we again use the Flory-Huggins model in the regime, where $\chi_{ij} \propto \gamma_{IJ}$. The final step is to adjust the volume fractions $\{\overline \phi_i\}$ for components, such that the volume fractions of separated phases are consistent with the target structure. This way were able to successfully construct the model parameters to produce target structures in simulations (see Fig.~\ref{fig4} and \href{http://www.princeton.edu/~akosmrlj/papers/phase_separation_design/videos/}{Video~S3}~\cite{video}). 

The graph theory approach presented in this paper is general. It can be applied to any model or experimental liquid mixture, and can also be generalized to other systems, such as block copolymers or liquid crystals.
In experiments it may be challenging to find immiscible fluids with sufficiently distinct surface energies to realize some complex target structures, but the promising new avenue is the phase separation of the solution of DNA strands~\cite{nguyen2017tuning,nguyen2019length}, where the interactions between DNA strands can be programmed via their sequences. Note that in a liquid environment separated phases continue to coarsen over time, but in some applications it may be beneficial to produce monodisperse structured droplets. Monodisperse structured droplets can be produced very efficiently with microfluidic devices~\cite{utada2005monodisperse,shah2008designer,choi2013one}. This can also be achieved by infusing a liquid mixture in a non-wetting elastomer, where the elastic deformation of the elastomer matrix can arrest the coarsening to produce monodisperse droplets~\cite{style2018liquid, kim2020extreme, rosowski2020elastic,kothari2020effect}. 
We hope that our study will stimulate further theoretical and experimental investigation of phase separation of multicomponent liquid mixtures in a wide range of fields, including biology (intracellular phase separation), chemical engineering (drugs and chemical microreactors), and environment (CO$_2$ sequestration and oil recovery).

\begin{acknowledgments}
This research was primarily supported by NSF through the Princeton
University’s Materials Research Science and Engineering Center
DMR-1420541 and through the REU Site EEC-1559973. We
would like to acknowledge useful discussions with Mikko Haataja, Yaofeng Zhong, Howard Stone, and Xiaoting Sun.
\end{acknowledgments}

\bibliography{library}

\end{document}


\title{Supplemental Material: Designing morphology of\\ separated phases in multicomponent liquid mixtures}
\author{Sheng Mao}
\affiliation{Department of Mechanics and Engineering Science, BIC-ESAT, College of Engineering, Peking University, Beijing 100871, People's Republic of China}
\affiliation{Department of Mechanical and Aerospace Engineering, Princeton University, Princeton, New Jersey 08544, USA}

\author{Milena S. Chakraverti-Wuerthwein}
\affiliation{Department of Physics, Princeton University, Princeton, New Jersey 08544, USA}

\author{Hunter Gaudio}
\affiliation{Department of Mechanical and Aerospace Engineering, Princeton University, Princeton, New Jersey 08544, USA}
\affiliation{Department of Mechanical Engineering, Villanova University, Villanova, Pennsylvania, 19085, USA}

\author{Andrej Ko\v{s}mrlj}
\affiliation{Department of Mechanical and Aerospace Engineering, Princeton University, Princeton, New Jersey 08544, USA}
\affiliation{Princeton Institute for the Science and Technology of Materials (PRISM),
Princeton University, Princeton, New Jersey 08544, USA}


\maketitle


\section{Simulation Methods}

Here, we briefly summarize numerical simulations, which are based on the code that was developed for our previous work~\cite{mao2019phase}.
The volume fraction fields $\{\phi_i(\boldsymbol{x})\}$ evolve via a so-called model B or Cahn--Hilliard dynamics~\cite{cahn1958, hohenberg1977theory}:
\begin{equation}
\frac{\partial \phi_i}{\partial t} =  \nabla \cdot \left[ \sum_{j} \tilde{M}_{ij}  \nabla \tilde{\mu}_j \right],
\label{eqn:governing-modelB}
\end{equation}
where we introduced the dimensionless chemical potentials $\tilde{\mu}_j =1 + \ln \phi_j + \sum_{k=1}^{N} \chi_{jk}  (1 + \lambda^2 \nabla^2) \phi_k$.  We adopted the Kramer's model~\cite{kramer1984interdiffusion} for the normalized Onsager mobility coefficients
$\tilde{M}_{ij}= D _{ij} \left( \phi_i \delta_{ij} - \phi_i \phi_j \right)$ to enforce the constraint $\sum_{i} \phi_i = 1$. When all components have identical diffusion coefficient $D_{ij} \equiv D$, then the Eq.~(\ref{eqn:governing-modelB}) can be re-written as
\begin{equation}
\frac{\partial \phi_i}{\partial t} =  D \nabla \cdot \left[ \phi_i \sum_{j}\left(\delta_{ij} - \phi_j \right)  \nabla \tilde{\mu}_j  \right].
\label{eqn:governing-phi}
\end{equation} 
Note that there are only $N-1$ independent volume fractions and $N-1$ independent chemical potentials. Note also that the interaction parameters $\{\chi_{ij}\}$ need to satisfy the condition $\sum_{i,j=1}^{N} a_i \chi_{ij} a_j <0$ for any $\lbrace a_i\rbrace$ with $\sum_{i=1}^{N} a_i=0$ to ensure the stability of interfaces~\cite{mao2019phase}.

The nonlinear partial differential equations in Eqn.~(\ref{eqn:governing-phi}) were solved numerically in a 3D cubic box with linear dimension $L$ discretized with $128 \times 128 \times 128$ uniform grid points and periodic boundary conditions.
A semi-implicit time-integration scheme \cite{zhu1999coarsening} was used, which enabled us to use relatively large time steps.  
To do so, we first discretized Eqn.~(\ref{eqn:governing-phi}) in time and separated the implicit linear and the explicit non-linear terms following the usual IMEX (implicit-explicit) scheme\cite{ascher1997implicit} as 
\begin{equation}
\label{eqn:governing-numerical}
\frac{\phi_i^{n+1} - \phi_i^{n}}{\Delta t} = N_i(\phi_i^n) + L_i(\phi_i^{n+1}), 
\end{equation}
where $\phi^{n}_i(\boldsymbol{x})$ is the volume fraction field of component $i$ at time step $n$. $N_i$ and $L_i$ denote the nonlinear and linear parts of the right hand side of Eqn.~(\ref{eqn:governing-phi}), respectively. Following the procedure in Ref.~\cite{zhu1999coarsening}, we introduced an artificial linear $\nabla^4$ term to stabilize the nonlinear term as 
\begin{eqnarray}
&& N_i(\{\phi_i\}) =  D \nabla \cdot \left[ \phi_i \sum_{j}\left(\delta_{ij} - \phi_j \right) \nabla \tilde{\mu}_j  \right] + A D \lambda^2 \nabla^4 \phi_i,  \\
&& L_i(\{\phi_i\}) =  - A D\lambda^2 \nabla^4 \phi_i, 
\end{eqnarray}
where the numerical prefactor $A=0.5\max \{\chi_{ij}\}$ is chosen empirically to ensure numerical stability. When evaluating nonlinear terms $N_i(\{\phi_i\})$, the products of composition fields $\phi_i^n (\boldsymbol{x})$ are carried out in real space, while the spatial derivatives are evaluated in Fourier representation $\hat{\phi}_{i}^{n} (\boldsymbol{ k})=\int_V  d \boldsymbol{x} \, e^{-i \boldsymbol{k} \cdot \boldsymbol{x}} \phi_i^n (\boldsymbol{x})/V$. The Fast Fourier Transform (FFT) algorithm was used to convert back and forth between real space and Fourier space representations~\cite{cooley1969fast}. In Fourier space, the implicit Eq.~(\ref{eqn:governing-numerical}) can be solved to obtain
\begin{equation}
 \hat{\phi}_{i}^{n+1} = \frac{ \hat{\phi_i^n} + \hat{N}_i(\phi^n_i)\Delta t }{1 + A\lambda^2k^4 D\Delta t},
\end{equation}
where $\hat{\cdot}$ denotes a Fourier transform and $k = |\boldsymbol{k}|$ is the magnitude of the wave vector $\boldsymbol{k}$.

To make equations dimensionless, the lengths are measured in units of the cubic box size $L$ and time is measured in the units of $\tau=\lambda^2/D$, which describes the characteristic time of diffusion across the interface between two phases. We chose $\lambda/L=0.45\times 10^{-2}$ and a time step $\Delta t = \tau/2$. For the initial conditions we set $\phi_i({\boldsymbol{x}}) = \bar{\phi}_i + \eta_i({\boldsymbol{x}})$, where $\eta_i({\boldsymbol{x}})$ is a uniform random noise with small magnitude and $0$ mean), and then the simulation runs for a total duration of $10^5-10^6 \tau$. The interaction parameters $\{\chi_{ij}\}$ and the average volume fractions $\{\bar \phi_i \}$ used in simulations are reported in Table~\ref{table:parameters}.

ParaView~\cite{ahrens2005paraview} was used for visualization, where we used isovolumes to indicate phases that are enriched in one of the components: red ($\phi_1 > \phi_\text{cutoff}$), green ($\phi_2 > \phi_\text{cutoff}$), blue ($\phi_3 > \phi_\text{cutoff}$), white ($\phi_4 > \phi_\text{cutoff}$), and dark gray ($\phi_5 > \phi_\text{cutoff}$). The threshold volume fraction for isovolumes was set to $\phi_\text{cutoff}=0.5-0.6$.

\begin{table}[h!]
 \centering
 \vspace{-0.3cm}
\caption{Simulation parameters
}
\label{table:parameters}
\def\arraystretch{1.15}
\begin{tabular}{|P{1.5cm}| P{8.0cm} | c|} 
\hline
Figure & interaction parameters & volume fractions\\
\hline
Fig.~1a & $\chi_{12}=\chi_{13}=\chi_{23}=3.25$ & $\{ \overline \phi_i\}=\{0.333, 0.333, 0.334\}$\\
Fig.~1a & $\chi_{12}=\chi_{13}=\chi_{23}=3.25$ & $\{ \overline \phi_i\}=\{0.15, 0.15, 0.70\}$\\
Fig.~1b & $\chi_{12}=\chi_{23}=2.5, \chi_{13}=5.5$ & $\{ \overline \phi_i\}=\{0.333, 0.333, 0.334\}$ \\
Fig.~1b & $\chi_{12}=\chi_{23}=2.5, \chi_{13}=5.5$ & $\{ \overline \phi_i\}=\{0.10, 0.20, 0.70\}$ \\
\hline
Fig.~2a & $\chi_{12}=\chi_{13}=\chi_{23}=\chi_{14}=\chi_{24}=\chi_{34}=5.0$ & $\{ \overline \phi_i\}=\{0.25, 0.25, 0.25, 0.25\}$ \\
Fig.~2b & $\chi_{12}=\chi_{13}=\chi_{23}=\chi_{14}=4.0$, $\chi_{24}=\chi_{34}=8.2$ & $\{ \overline \phi_i\}=\{0.25, 0.25, 0.25, 0.25\}$ \\
Fig.~2c & $\chi_{12}=\chi_{13}=\chi_{23}=\chi_{14}=4.0$, $\chi_{24}=8.5$, $\chi_{34}=13.0$ & $\{ \overline \phi_i\}=\{0.25, 0.25, 0.25, 0.25\}$ \\
\hline
Fig.~3a & $\chi_{12}=\chi_{13}=\chi_{23}=\chi_{14}=\chi_{24}=\chi_{34}=5.0$ & $\{ \overline \phi_i\}=\{0.25, 0.25, 0.25, 0.25\}$ \\
Fig.~3a & $\chi_{12}=\chi_{13}=\chi_{23}=\chi_{14}=\chi_{24}=\chi_{34}=5.0$ & $\{ \overline \phi_i\}=\{0.10, 0.10, 0.10, 0.70\}$ \\
Fig.~3b & $\chi_{12}=\chi_{13}=\chi_{23}=\chi_{14}=\chi_{34}=4.5$, $\chi_{24}=10.0$ & $\{ \overline \phi_i\}=\{0.25, 0.25, 0.25, 0.25\}$ \\
Fig.~3b & $\chi_{12}=\chi_{13}=\chi_{23}=\chi_{14}=\chi_{34}=4.5$, $\chi_{24}=10.0$ & $\{ \overline \phi_i\}=\{0.12, 0.06, 0.12, 0.70\}$ \\
Fig.~3c & $\chi_{12}=\chi_{13}=\chi_{23}=\chi_{14}=4.0$, $\chi_{24}=\chi_{34}=8.2$ & $\{ \overline \phi_i\}=\{0.25, 0.25, 0.25, 0.25\}$ \\
Fig.~3c & $\chi_{12}=\chi_{13}=\chi_{23}=\chi_{14}=4.0$, $\chi_{24}=\chi_{34}=8.2$ & $\{ \overline \phi_i\}=\{0.20, 0.05, 0.05, 0.70\}$ \\
Fig.~3d & $\chi_{12}=\chi_{13}=\chi_{23}=7.5$,
$\chi_{14}=\chi_{24}=\chi_{34}=3.5$ & $\{ \overline \phi_i\}=\{0.25, 0.25, 0.25, 0.25\}$ \\
Fig.~3d & $\chi_{12}=\chi_{13}=\chi_{23}=10.5$,
$\chi_{14}=\chi_{24}=\chi_{34}=5.0$ & $\{ \overline \phi_i\}=\{0.10, 0.10, 0.10, 0.70\}$ \\
Fig.~3e & $\chi_{12}=\chi_{23}=\chi_{14}=\chi_{34}=3.0$, $\chi_{13}=\chi_{24}=6.5$ & $\{ \overline \phi_i\}=\{0.25, 0.25, 0.25, 0.25\}$ \\
Fig.~3e & $\chi_{12}=\chi_{23}=\chi_{14}=\chi_{34}=3.0$, $\chi_{13}=\chi_{24}=6.5$ & $\{ \overline \phi_i\}=\{0.10, 0.10, 0.10, 0.70\}$ \\
Fig.~3f & $\chi_{12}=\chi_{23}=\chi_{34}=2.5$, $\chi_{13}=5.5$, $\chi_{14}=8.5$, $\chi_{24}=6.5$ & $\{ \overline \phi_i\}=\{0.25, 0.25, 0.25, 0.25\}$ \\
Fig.~3f & $\chi_{12}=\chi_{23}=\chi_{34}=3.3$, $\chi_{13}=6.9$, $\chi_{14}=10.5$, $\chi_{24}=7.5$ & $\{ \overline \phi_i\}=\{0.03, 0.07, 0.20, 0.70\}$ \\
\hline
\multirow{2}{*}{Fig.~4a} & $\chi_{12}=\chi_{23}=\chi_{45}=8.2$,  $\chi_{13}=\chi_{14}=4.0$, & \multirow{2}{*}{$\{ \overline \phi_i\}=\{0.11,0.11,0.11,0.62,0.05\}$} \\
& $\chi_{24}=\chi_{34}=\chi_{15}=\chi_{25}=\chi_{35}=4.0$ & \\
\multirow{2}{*}{Fig.~4b} & $\chi_{12}=\chi_{13}=\chi_{14}=\chi_{24}=\chi_{34}=\chi_{35}=4.0$, & \multirow{2}{*}{$\{ \overline \phi_i\}=\{0.11,0.11,0.08,0.65,0.05\}$} \\
& $\chi_{23}=\chi_{15}=\chi_{45}=8.2$, $\chi_{25}=12.4$ & \\
\hline
Fig.~S1a & $\chi_{12}=\chi_{13}=\chi_{23}=\chi_{14}=\chi_{24}=\chi_{34}=5.0$ & $\{ \overline \phi_i\}=\{0.25, 0.25, 0.25, 0.25\}$ \\
Fig.~S1b.1 & $\chi_{12}=\chi_{13}=\chi_{23}=\chi_{34}=4.0$, $\chi_{14}=6.0$, $\chi_{24}=9.0$  & $\{ \overline \phi_i\}=\{0.25, 0.25, 0.25, 0.25\}$ \\
Fig.~S1b.2 & $\chi_{12}=\chi_{13}=\chi_{23}=\chi_{14}=\chi_{34}=4.5$, $\chi_{24}=10.0$ & $\{ \overline \phi_i\}=\{0.25, 0.25, 0.25, 0.25\}$ \\
Fig.~S1c.1 & $\chi_{12}=\chi_{13}=\chi_{23}=\chi_{14}=4.0$, $\chi_{24}=\chi_{34}=8.2$ & $\{ \overline \phi_i\}=\{0.25, 0.25, 0.25, 0.25\}$ \\
Fig.~S1c.2 & $\chi_{12}=\chi_{13}=\chi_{23}=\chi_{14}=4.0$, $\chi_{24}=8.5$, $\chi_{34}=13.0$ & $\{ \overline \phi_i\}=\{0.25, 0.25, 0.25, 0.25\}$ \\
Fig.~S1d.1 & $\chi_{12}=\chi_{13}=\chi_{23}=7.5$,
$\chi_{14}=\chi_{24}=\chi_{34}=3.5$ & $\{ \overline \phi_i\}=\{0.25, 0.25, 0.25, 0.25\}$ \\
Fig.~S1d.2 & $\chi_{12}=\chi_{13}=5.2$, $\chi_{23}=10.6$,
$\chi_{14}=\chi_{24}=\chi_{34}=2.5$ & $\{ \overline \phi_i\}=\{0.25, 0.25, 0.25, 0.25\}$ \\
Fig.~S1e.1 & $\chi_{12}=\chi_{23}=\chi_{34}=3.0$
$\chi_{13}=\chi_{24}=6.5$, $\chi_{14}=4.0$ & $\{ \overline \phi_i\}=\{0.25, 0.25, 0.25, 0.25\}$ \\
Fig.~S1e.2 & $\chi_{12}=\chi_{14}=3.2$, $\chi_{13}=6.4$, $\chi_{23}=\chi_{34}=3.0$, $\chi_{24}=6.2$ & $\{ \overline \phi_i\}=\{0.25, 0.25, 0.25, 0.25\}$ \\
Fig.~S1e.3 & $\chi_{12}=\chi_{23}=\chi_{14}=\chi_{34}=3.0$, $\chi_{13}=\chi_{24}=6.5$ & $\{ \overline \phi_i\}=\{0.25, 0.25, 0.25, 0.25\}$ \\
Fig.~S1f.1 & $\chi_{12}=\chi_{23}=\chi_{34}=2.5$, $\chi_{13}=5.5$, $\chi_{14}=8.5$, $\chi_{24}=6.5$ & $\{ \overline \phi_i\}=\{0.25, 0.25, 0.25, 0.25\}$ \\
Fig.~S1f.2 & $\chi_{12}=\chi_{23}=\chi_{34}=2.5$, $\chi_{13}=\chi_{24}=5.1$, $\chi_{14}=7.7$  & $\{ \overline \phi_i\}=\{0.25, 0.25, 0.25, 0.25\}$ \\
Fig.~S1f.3 & $\chi_{12}=\chi_{23}=\chi_{34}=2.2$, $\chi_{13}=4.6$, $\chi_{14}=7.0$, $\chi_{24}=9.4$ & $\{ \overline \phi_i\}=\{0.25, 0.25, 0.25, 0.25\}$ \\
\hline            
\end{tabular}
\end{table}

\clearpage

\begin{figure*}[h!]
\includegraphics[width=\linewidth]{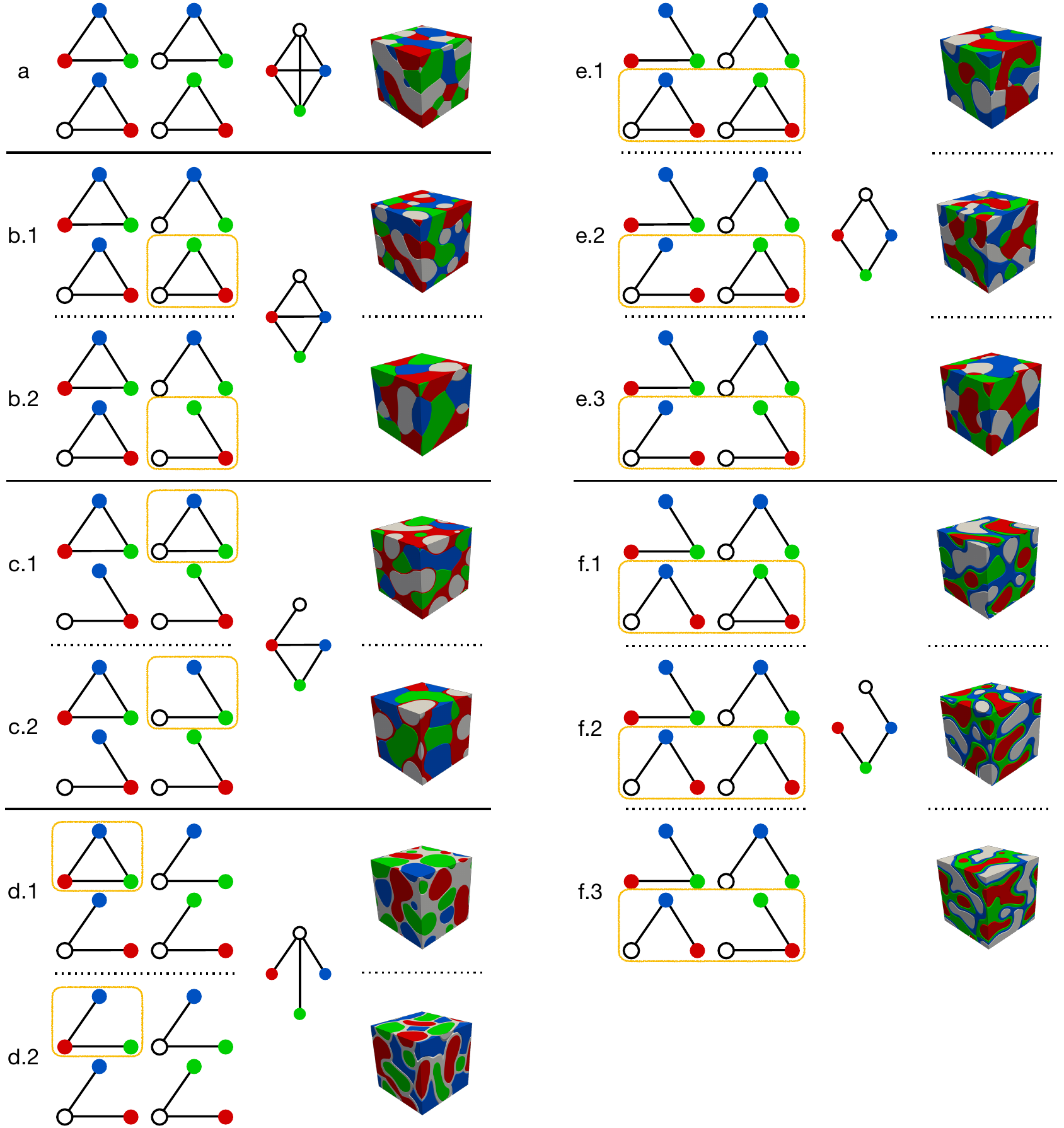}
\caption{\label{fig:4comp_all_solutions} 
All distinct sets of the wetting conditions ($4$ graphs of triplets) that can be realized for a mixture with $N_p=4$ phases. These sets are grouped according to the connectivity graphs with $4$ vertices describing the topology of separated phases. For each group the yellow boxes indicate the wetting conditions that differ between sets. For each set we show a simulation snapshot at $10^6$ timesteps. The simulation parameters are given in Table~\ref{table:parameters}.  
}
\end{figure*}

\begin{figure*}[h!]
\includegraphics[width=\linewidth]{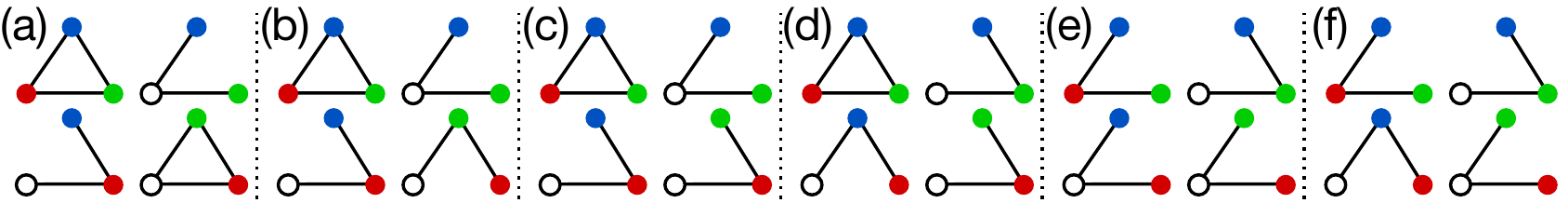}
\caption{\label{fig:4comp_all_no-sol} 
All distinct sets of the wetting conditions ($4$ graphs of triplets) that cannot be realized for a mixture with $N_p=4$ phases. 
}
\end{figure*}

\begin{figure*}[h!]
\includegraphics[width=\linewidth]{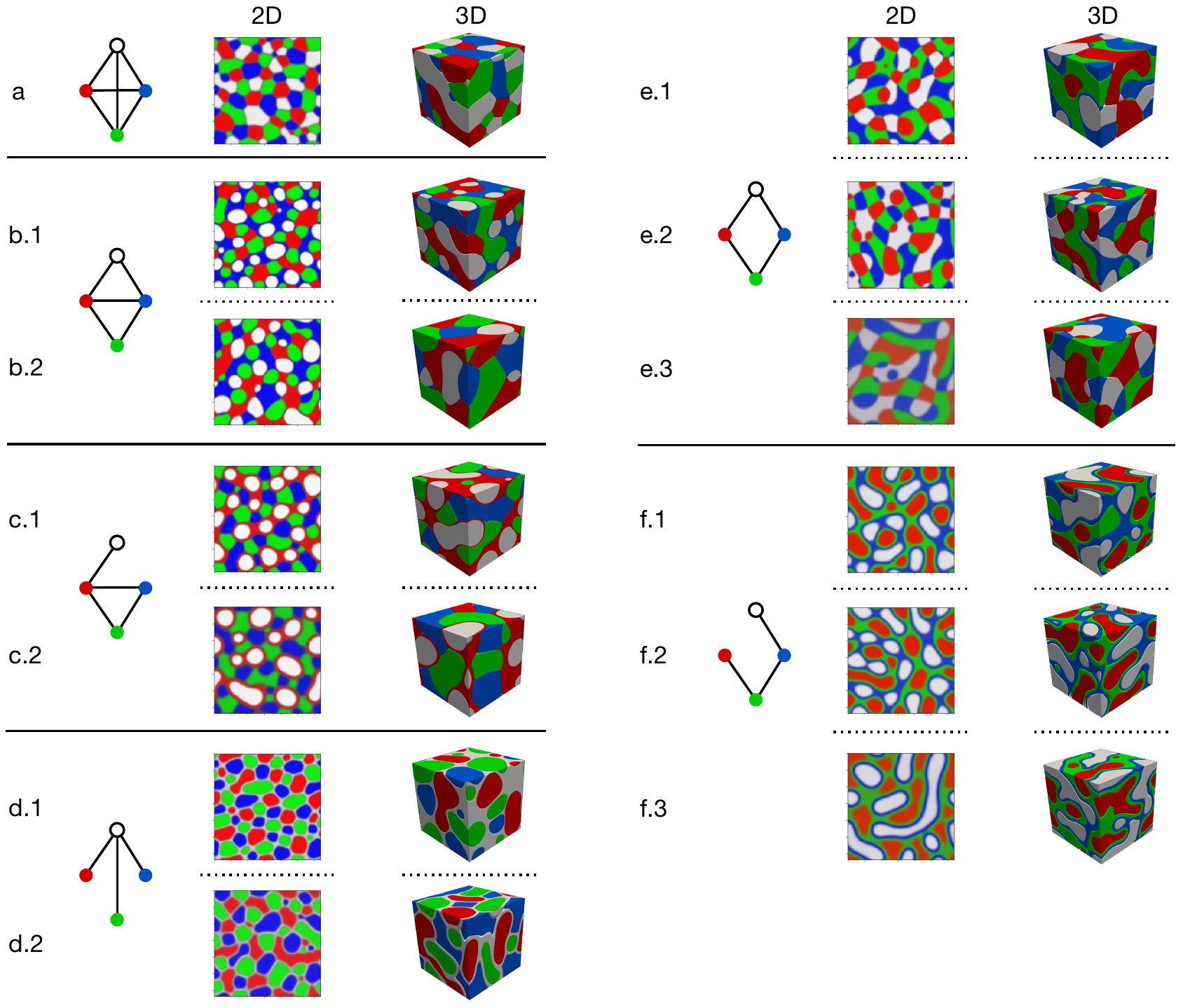}
\caption{
Comparison of simulation snapshots in 2D and 3D for all distinct sets of the wetting conditions presented in Fig.~\ref{fig:4comp_all_solutions}. The simulation parameters are identical for 2D and 3D simulations and they are the same as in Fig.~\ref{fig:4comp_all_solutions}. 
}
\end{figure*}

\begin{figure*}[t!]
\includegraphics[scale=1]{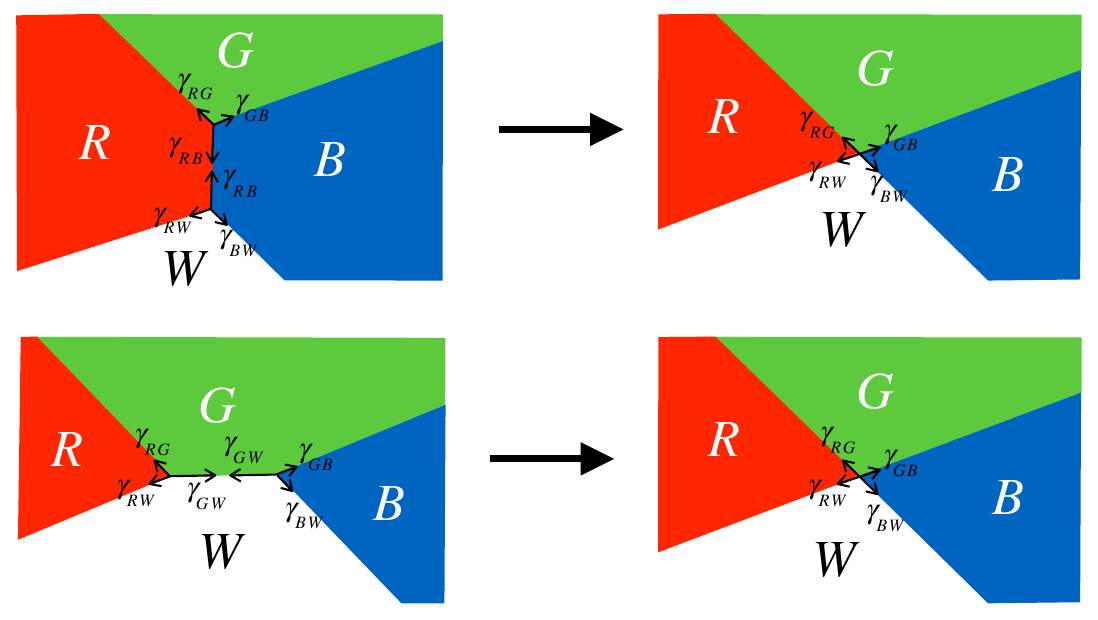}
\caption{\label{fig:4-junction} 
Stability of quadruple junctions. For simulation snapshots in Fig.~\ref{fig:4comp_all_solutions}e the quadruple junctions are stable, because triple junctions get pulled together due to the force imbalance of surface tensions. For the case in Fig.~\ref{fig:4comp_all_solutions}e.1, the $R$-$G$-$B$ junction gets pulled ($\gamma_{RB}>\gamma_{RG}+\gamma_{GB}$) toward the stable $R$-$B$-$W$ junction and the $G$-$B$-$W$ junction gets pulled ($\gamma_{GW}>\gamma_{GB}+\gamma_{BW}$) toward the stable $R$-$G$-$W$ junction. For the case in Fig.~\ref{fig:4comp_all_solutions}e.2, the $R$-$G$-$B$ and $R$-$B$-$W$ junctions are pulled toward each other ($\gamma_{RB}>\gamma_{RG}+\gamma_{GB}$, $\gamma_{RB}>\gamma_{RW}+\gamma_{RG}$)  and the $G$-$B$-$W$ junction gets pulled ($\gamma_{GW}>\gamma_{GB}+\gamma_{BW}$) toward the stable $R$-$G$-$W$ junction.
For the case in Fig.~\ref{fig:4comp_all_solutions}e.3, the $R$-$G$-$B$ and $R$-$B$-$W$ junctions are pulled toward each other ($\gamma_{RB}>\gamma_{RG}+\gamma_{GB}$, $\gamma_{RB}>\gamma_{RW}+\gamma_{RG}$)  and the $G$-$B$-$W$ and $R$-$G$-$W$ junctions are pulled toward each other ($\gamma_{GW}>\gamma_{GB}+\gamma_{BW}$, $\gamma_{GW}>\gamma_{GR}+\gamma_{RW}$).
}
\end{figure*}

\bibliography{library}